\documentclass[a4paper,fleqn]{cas-sc}

\usepackage[numbers]{natbib}

\usepackage{subcaption}

\def\tsc#1{\csdef{#1}{\textsc{\lowercase{#1}}\xspace}}
\tsc{WGM}
\tsc{QE}

\begin{document}
\let\WriteBookmarks\relax
\def\floatpagepagefraction{1}
\def\textpagefraction{.001}

\shorttitle{Lunar ISRU habitat based on TIA}    

\shortauthors{L. Schnelle, K.-U. Schr\"oder, A.C. Niemeyer, Y. Estrin}  

\title [mode = title]{A pyramidal ISRU lunar habitat design based on topological interlocking of sintered regolith blocks}  



%

\author[1, 2]{Lukas Schnelle}[orcid=0009-0009-6998-7433,]

\cormark[1]


\ead{lukas.schnelle@sla.rwth-aachen.de}


\credit{Conceptualization, Methodology, Formal analysis, Investigation, Writing - Original Draft, Visualization}

\affiliation[1]{organization={Institute of Structural Mechanics and Lightweight Design, RWTH Aachen University},
            addressline={W\"ullnerstraße 7}, 
            city={Aachen},
            citysep={},
            postcode={52062}, 
            country={Germany}}

\author[1]{Kai-Uwe Schr\"oder}[orcid=0000-0002-3171-962X]

\credit{Conceptualization, Validation, Writing - Review \& Editing, Supervision}

\affiliation[2]{organization={Chair of Algebra and Representation Theory, RWTH Aachen University},
            addressline={Pontdriesch 10-16}, 
            city={Aachen},
            citysep={},
            postcode={52062}, 
            country={Germany}}

\author[2]{Alice C. Niemeyer}[orcid=0000-0002-2163-3240]
\credit{Conceptualization, Validation, Writing - Review \& Editing, Supervision}

\author[3]{Yuri Estrin}[orcid=0000-0001-7784-5704]
\credit{Conceptualization, Validation, Writing - Review \& Editing, Supervision}

\affiliation[3]{organization={Department of Materials Science and Engineering, Monash University},
            addressline={Building 37, 19 Rainforest Walk}, 
            city={Clayton},
            citysep={},
            state={VIC},
            postcode={3800}, 
            country={Australia}}

\cortext[1]{Corresponding author}



\begin{abstract}
In this framing study, we propose a new lunar‑habitat design based on topological interlocking of the building blocks. The design combines block geometries that permit interlocking without binders or connectors with a pyramidal overall structure. This approach is well suited for ISRU‑enabled extraterrestrial construction. Finite element simulations employing the Drucker-Prager model confirm that habitats manufactured from lunar regolith using this concept are tolerant to local failures or missing blocks. We further suggest that this approach can be extended and improved by optimising the geometry of the interlockable building blocks. \end{abstract}



\begin{keywords}
 ISRU \sep lunar exploration \sep regolith \sep topological interlocking
\end{keywords}

\maketitle

\section{Introduction to the design of lunar habitats}\label{sec:intro}
Across antiquity, the Middle Ages, and into modern times, thinkers and writers have envisioned the possibility of travelling to the Moon. Generations of readers were fascinated by books about space travel written in the 19th and 20th centuries by such visionaries as Jules Verne, H.G. Wells, Robert A. Heinlein and Isaac Asimov. What once seemed entirely utopian became reality when Neil Armstrong set foot on the lunar surface during the historic Apollo~11 mission in 1969. With advances in space‑travel technology, the hope for establishing a presence of humans on the Moon, and potentially on Mars and other celestial bodies, gained momentum. Space agencies in various countries are currently developing large-scale programs targeting future colonisation of the Moon, such as the Artemis program of NASA, ESA’s Moon Village Concept, and International Lunar Research Station project run jointly by China and Russia \cite{evans2020, heinicke2021, xu2022}. The prospects of establishing a presence of humans on the Moon have also become attractive commercially, and a number of private companies including SpaceX and Blue Origin and others \cite{lueders2021} have embarked on lunar colonisation projects. 

In the pursuit of lunar colonisation, building habitats for humans is essential for ensuring their long-term survival. Even though the distance to the Moon is relatively small on the cosmic scale, transporting materials required for construction is prohibitively expensive and logistically impractical. This is the reason why recent efforts have been directed towards in situ resource utilisation (ISRU) on the Moon. The material readily available on the lunar surface, lunar regolith, is at the centre of attention of space agencies and researchers \cite{mendell1999}. This material occurs in the form of fine dust that can be sintered into building blocks \cite{bao2024}. In terms of density, mechanical properties and manufacturability, the resulting material is comparable with a porous ceramic, although sintered regolith has greater variability in its properties depending on the local lunar soil composition. In laboratory experiments, sintered regolith simulants have been shown to withstand lunar‑surface environmental conditions \cite{ginespalomares2023}. The energy required for sintering can be provided by the readily available sunlight. The use of this energy source to manufacture regolith-based construction components has been extensively explored in the literature \cite{fateri2019, meurisse2018, suhaizan2023}.

A general consensus is that extraterrestrial construction must be modular. Not only does modularity enable progressive expansion as missions grow, but it also facilitates the use of robotics for autonomous assembly \cite{caluk2023}. It is also conducive to module recycling, enabling reconfigurable structures. In addition, modular construction helps mitigate the inherent brittleness of sintered regolith \cite{khor2004}. For effective use of this design concept it is desirable to minimise the number of block variants. In addition, the individual blocks should be sized for easy handling. A common drawback of the building blocks used for modular construction in terrestrial applications \cite{block2006b, block2006a} is the need for adhesives, mortar, or connectors. The very low atmospheric pressure on the Moon makes curing of adhesives or mortar problematic, while the scarcity of water hinders mortar production. Additionally, any adhesive connection has inherent uncertainties, as a specific adhesive joint cannot be tested non-destructively. The use of connectors is also detrimental to the structural integrity of assemblies due to stress concentrations they induce. 

A viable alternative to common geometries of the building blocks are designs based on the concept of topological interlocking first suggested in \cite{dyskin2001, dyskin2002}. This design methodology was adopted and further extended by several authors, notably from the groups of Thomas Siegmund and Francois Barthelat \cite{siegmund2016}, Vinayak Krishnamurthy and Ergun Akleman \cite{estrin2021}, Yi Min Xie \cite{wang2025b, wang2026}, Alice Niemeyer \cite{goertzen2022, goertzen2025}, and most recently, Chiara Daraio \cite{lu2024}. Topological interlocking structures are assemblies of building blocks that hold together through their geometry and mutual arrangement, rather than through binders or connectors. Their use in extraterrestrial construction was first proposed in an early paper \cite{dyskin2005}, which suggested using solar energy to sinter the sun’s energy for sintering regolith into topologically interlockable blocks. Although there are quite a few proposed building block geometries that permit topological interlocking assembly, the majority of the reported solutions focus on planar block arrangements. Human habitats require a substantial amount of volume \cite{landon2019}, which cannot be created using elements derived exclusively from  two‑dimensional assemblies. That is why it is important to find a three‑dimensional habitat geometry that can be realized using a low number of identical, truly ``one‑size‑for‑all”, topologically interlocking blocks. 

One of the well known layouts for lightweight structures subjected to internal pressure is a doubly curved shell, often realised as a dome or hemisphere. These funicular (compression-only) structures carry loads through compression thereby avoiding bending and enhancing structural robustness \cite{allen2009}. A radically different, kirigami-based, design for lunar habitat was developed by Chiara Daraio and her team at Caltech who proposed planar non-periodic metamaterials for producing load-bearing pop-up 3D structures, see their recent publication \cite{niloy2025}. 

If one chooses to adopt more traditional modular designs - the approach we advocate - an inherent challenge arises: constructing a tessellated double-curvature shell with adaptable size requires constituent blocks that vary slightly in shape. \cite[cf. fig. 2 (d, e)]{liu2022}. Specifically, mapping a planar assembly onto a hemisphere results in non‑uniform distortions of individual blocks (Theorema egregium (Gauss) \cite{do2016}). In search of an alternative to a hemisphere, we were guided by the following consideration: An atmosphere inside the habitat exerts pressure normal to the enclosing surfaces, while gravity always acts downward. These two forces should ideally balance, thereby reducing the structural loads. The problem is compounded by the reduced gravity on the Moon, suggesting that vertical walls should not be employed. 

To resolve these issues we propose a lunar habitat geometry based on a pyramidal structural form. The faces of a pyramid are equilateral triangles tessellated into monolayers of identical blocks that are interlocked topologically. In particular, any displacement of a block in the direction normal to the layer is kinematically constrained by its neighbouring blocks. We designed the shape of these interlockable blocks specifically for this application. The choice of pyramidal construction was also motivated by the millenia-long architectural tradition as evident both in the ancient Egyptian pyramids and in modern structures, such as the Louvre Pyramid, whose structural stability and aesthetic appeal are undisputable.

In the following sections, we present the details of the proposed hollow pyramid structures and demonstrate that they are mechanically viable in the lunar environment. This is shown using finite‑element (FE) simulations that account for block damage. The advantages of lunar structures composed of topologically interlocked regolith blocks, especially their tolerance to local failures, will also be discussed. 

\section{Geometry of the building blocks and arrangement into a pyramid}\label{sec:geom}

\subsection{Design considerations for the overall structure}\label{ssec:design-considerations}
As mentioned above, the main objective is to create a fully enclosed, pressurisable volume that provides a controlled environment in which astronauts can live and work. Pressurisation also allows the use of standard equipment without additional shielding, yielding further savings in the material that must be transported to the lunar surface. The design of the habitat is guided by the interplay of two principal loads: the internal pressure required for human habitation and the gravitational loading imposed by the lunar environment. 

The direction of the load induced by the internal pressure depends on the specific design of the habitat, as it always acts perpendicular to the enclosing surfaces. The load induced by gravity always acts downward. Ideally, these two loads would counteract each other and thus have no net effect on the overall structure. 
One way to achieve such a planar configuration with usable volume is to add a plate on top of vertical walls, as is commonly done on Earth. This configuration is unfavourable, however, since the load exerted on a vertical wall by the internal pressure contains no upward component, the structure must carry this load without any compensation by the gravitational load. 

An approach frequently used in structural engineering when the internal pressure of a structure is the dominating load case is a shell, as in this particular case shells react only with membrane forces, not bending. Especially hemispheres are desirable because they have the same curvature everywhere and thus can support an internal pressure load well. To apply the tiling techniques developed in \cite{goertzen2022, goertzen2024}, one begins with a tiling of the sphere. Since the aim is to construct interchangeable tiles, the individual elements must be congruent. The classification of tilings of a sphere shows that only the Platonic solids meet this requirement \cite{agaoka2002}. Even when the condition is relaxed to allow a small number of congruency classes, the resulting tilings still admit only geometries with a fixed number of tiles. This predetermined tile count fixes the tile angles and thereby constrains the overall scale of the habitat. Crucially, it precludes any possibility of expansion: habitats constructed in this manner cannot be expanded using the same blocks. This severely limits the practical applicability of the method for deployable or evolving habitat systems.

Another conceivable approach would be to approximate a sphere using a polyhedron with a large number of faces. One example is provided by the well-known geodesic domes, frequently used in architecture. Mathematically, a geodesic dome is a polyhedron that arises as the dual of a Goldberg polyhedron, first introduced in \cite{goldberg1937}. Goldberg polyhedra are constructed by starting from a regular dodecahedron and inserting hexagons between its pentagons; iterating this procedure yields polyhedra with an increasingly large number of faces. They can be used in two distinct ways: one may allow a small number of different triangles (by their face‑congruency classes), which reintroduces the limitations on scalability described earlier, or one may fix the edge length of all faces, in which case the resulting shape gradually approaches an icosahedral geometry.

To avoid a ``frozen" configuration and ensure the expandability of the habitat, we impose the condition that all edge lengths of the tiles are the same. To that end, we introduce an icosahedron-based habitat geometry. By considering only one half of the icosahedron we obtain a pyramid with a pentagonal base. There are two more possible non-planar configurations for pyramids with equilateral triangles as sides: triangular and square-shaped bases. To minimize the non-functional volume, we focus on the square-shaped base, as this offers a viable compromise between the excessive height of a triangular based pyramid and the large amount of unusable, low-clearance areas of a pentagonal based pyramid. For simplicity, we consider only equilateral inclined faces of the pyramid, as this provides greater symmetry in the components that must be fabricated. 

The goal is to enclose a space by the faces of a pyramid with a sufficiently large usable base area while maintaining a reasonable size of the overall structure. We selected a square-shaped base with an edge length of approximately $7.5 \, \mathrm{m}$. One of the advantages of a pyramid with inclined faces formed by equilateral triangles is that with an appropriate tile geometry, scaled versions of the same pyramid can be assembled simply by adding or removing blocks. This means that the same tiles can be assemble an expanded (or contracted) pyramids allowing for different sizes depending on the current need.

One of the main advantages we see in constructing a lunar habitat based on a topological interlocking, is that the structure is redundant in the sense that a missing block does not directly result in a total loss of structural integrity, a behaviour already established for planar assemblies \cite{khor2004}. If, for example, a micrometeorite or a piece of equipment impacts the habitat, an individual block may be completely lost, yet the overall structure will withstand such an event. To ensure gas-tightness, a gas-tight membrane on the inside of the habitat can be used. 

\subsection{The individual block}\label{ssec:individual-block}
In this section we describe the geometry of an individual block (tile) that can be used to assemble a side of the pyramid, satisfying the requirements outlined above. For these blocks we adopt a maximum dimension of $1 \, \mathrm{m}$ in any direction to ensure manageable handling. Due to the specific interlocking construction we are considering here \cite{goertzen2024}, the scaling can be adapted to different fabrication setups or to the requirements of a specific assembly robot, if needed for a given application. 

Because the pyramid has equilateral triangles as its inclined faces, we begin with a smaller equilateral triangle $F$ as the basic unit for tiling a triangular face. By translating and rotating this smaller triangle $F$ in two directions, a planar tiling is obtained. From a group‑theoretical perspective, the triangle $F$ can be regarded as a fundamental domain of the crystallographic group $p6$. This group generates a tiling of the entire two‑dimensional plane from a single equilateral triangle through three‑fold rotations about each vertex and the associated transformations that produce hexagonal arrangements filling the plane. To obtain a larger tiled triangle, we select part of the plane. By adjusting the choice of this part, we therefore can control how finely subdivided the big triangle is.
Additionally, this allows not just the expansion of the structure we propose but also a reconfiguration to limitations of manufacturing or assembly if these should arise. In particular, if the equipment is only able to handle smaller parts than the $1 \, \mathrm{m}$ we assumed, scaling everything down to the required size while considering a finer tiling results in a structure of the same size. This is particularly important, as many aspects of manufacturing and assembly on the lunar surface are still unknown.

\begin{figure}[pos=ht]
    \centering
    \includegraphics[width=0.7\textwidth]{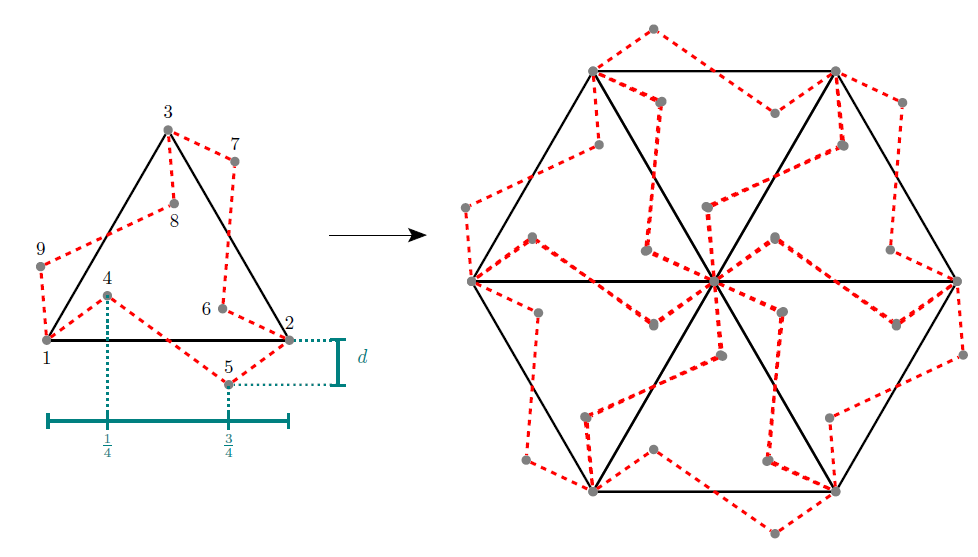}
    \caption{Procedure for generating a Starfish block: (a) Schematic view of the deformation of the equilateral triangle $F$ as a fundamental domain with the red lines representing the modified configuration. The grey dots are the vertices spanning $F$ as well as the newly introduced ones during deformation (b) The resulting tiling of the initial and modified fundamental domains. From \cite[Fig. 1]{schnelle2026}.}
    \label{fig:fund-dom}
\end{figure}

\begin{figure}[pos=ht]
    \centering
    \hfill
    \begin{subfigure}[b]{0.3\textwidth}
        \includegraphics[width=\textwidth]{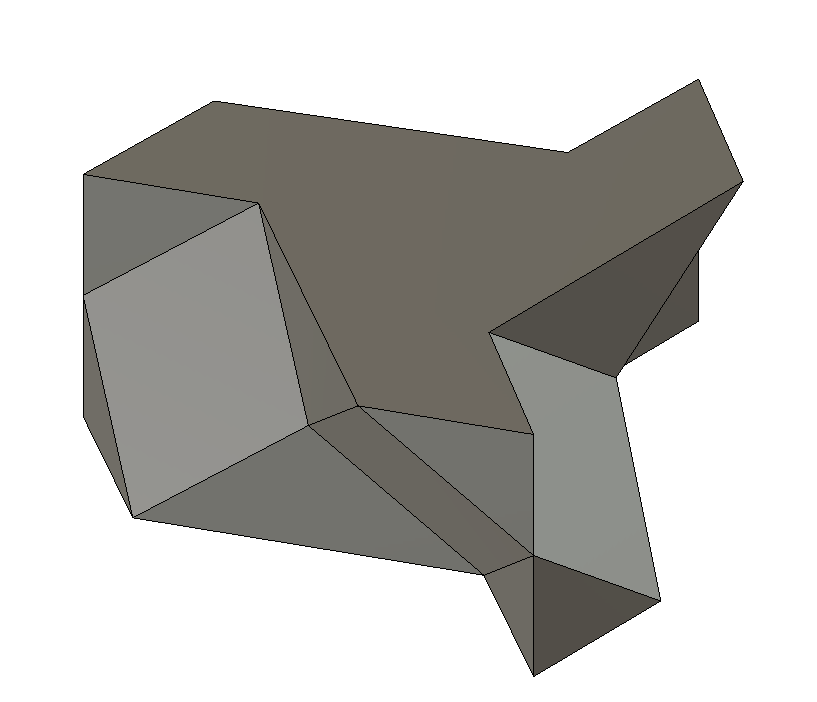}
    \end{subfigure}
    \hfill
    \begin{subfigure}[b]{0.3\textwidth}
        \includegraphics[width=\textwidth]{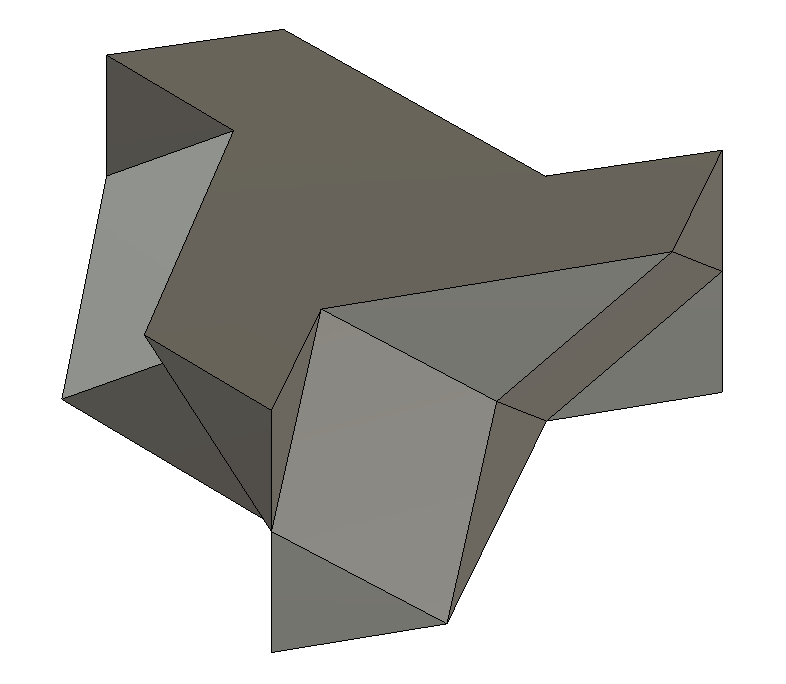}
    \end{subfigure}
    \hfill
    \begin{subfigure}[b]{0.3\textwidth}
        \includegraphics[width=\textwidth]{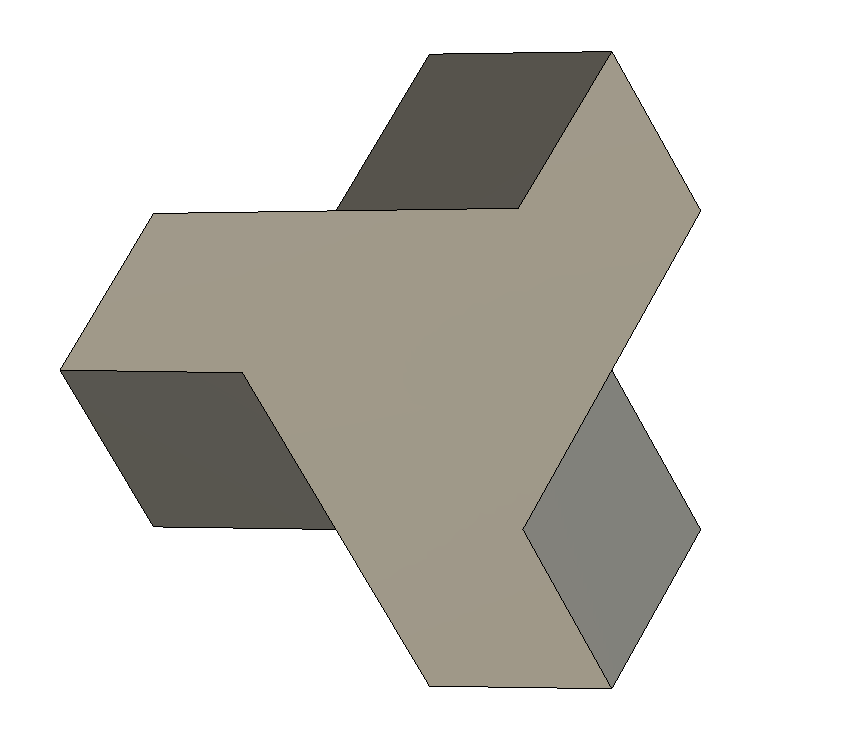}
    \end{subfigure}
    \hfill
    \caption{Different views of the individual double-Starfish block}
    \label{fig:individual-block}
\end{figure}

As discussed in \cite{goertzen2022, goertzen2024, schnelle2026}, an interlocking geometry in the plane can be generated from this fundamental domain by introducing a second, modified fundamental domain of the same crystallographic group. Here the modification is produced by subdividing each side of the small equilateral triangle in four equal segments by adding vertices at the subdivision points. Of these new vertices, the outer two are then displaced by $120 \, \mathrm{mm}$ normal to the original edge, one inward and one outward. To ensure that the resulting shape remains a fundamental domain - specifically, that it tiles the plane without gaps or self‑intersections under the same $p6$ symmetry - this modification must be applied consistently. Adjusting only one side would lead to gaps and overlaps between copies of the shape. However, performing the same modification on all three sides of the triangle, alternating the inward and outward displacements, produces a new tiling of the plane governed by the $p6$ crystallographic group. The modification described and the resulting tiling can be seen in Figure \ref{fig:fund-dom}.
We then place the original fundamental domain and the modified one on two parallel planes, here offset by $200 \, \mathrm{mm}$. As both fundamental domains tile their respective planes under the same $p6$ symmetry, interpolating between them yields a three dimensional block that fills the region between the two planes without gaps or intersections. In addition, we introduce a second modified tiling on a third offset plane, also offset by $200 \, \mathrm{mm}$, so that three $p6$ tilings lie parallel to one another at equal spacing. This second modified fundamental domain reverses the inward/outward assignment of the displaced vertices. Interpolating between each pair of tilings produces a block that fills the full $400 \, \mathrm{mm}$ thickness without intersections or voids. The resulting single block, shown in Figure \ref{fig:individual-block} from multiple viewpoints is referred to as the \emph{double-Starfish block}, since the interpolation between the fundamental domain and the singly modified domain is known as the Starfish block \cite{schnelle2026}. 

Every known instance of the method from \cite{goertzen2022, goertzen2024, schnelle2026}, to the authors’ knowledge, has produced a topological interlocking geometry, and we therefore assume that this construction likewise yields a topological interlocking assembly. This assumption is supported by the results reported in \cite{schnelle2026}, where different mechanical configurations of the double-Starfish block were tested and all exhibited positive stiffness.

One high-performing assemblage of double-Starfish blocks \cite{schnelle2026} has a $6 \, \mathrm{mm}$ configuration, which we therefore consider here as well. However, at the larger scale considered here, this translates to a deformation of $120 \, \mathrm{mm}$ in our geometry.

\subsection{Edges and apex}\label{ssec:edges-apex}
\begin{figure}[pos=ht]
    \centering
    \hfill
    \begin{subfigure}[b]{0.4\textwidth}
        \includegraphics[width=\textwidth]{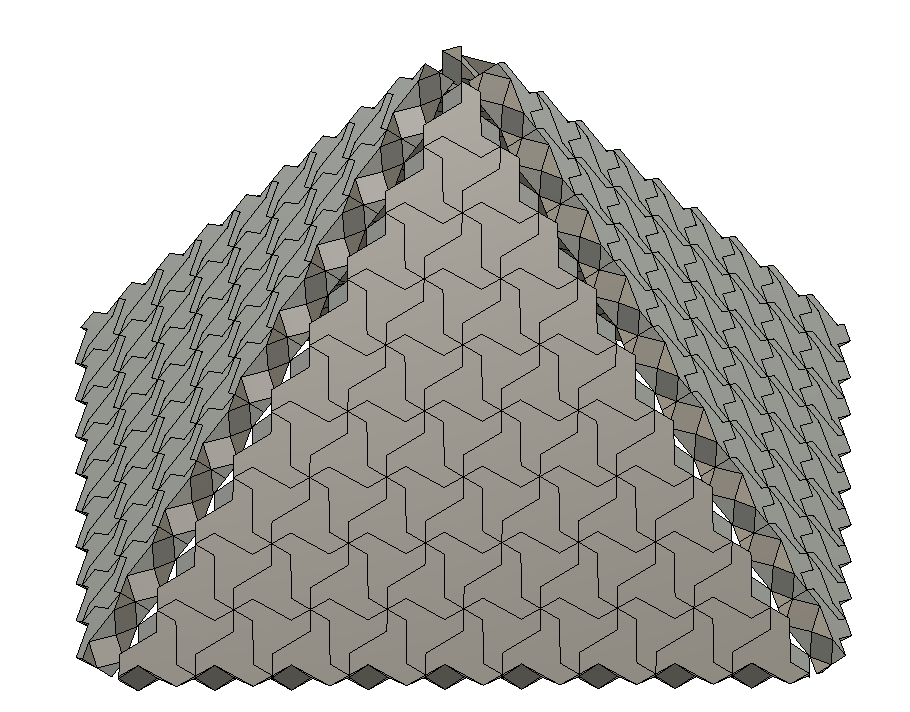}
    \end{subfigure}
    \hfill
    \begin{subfigure}[b]{0.49\textwidth}
        \includegraphics[width=\textwidth]{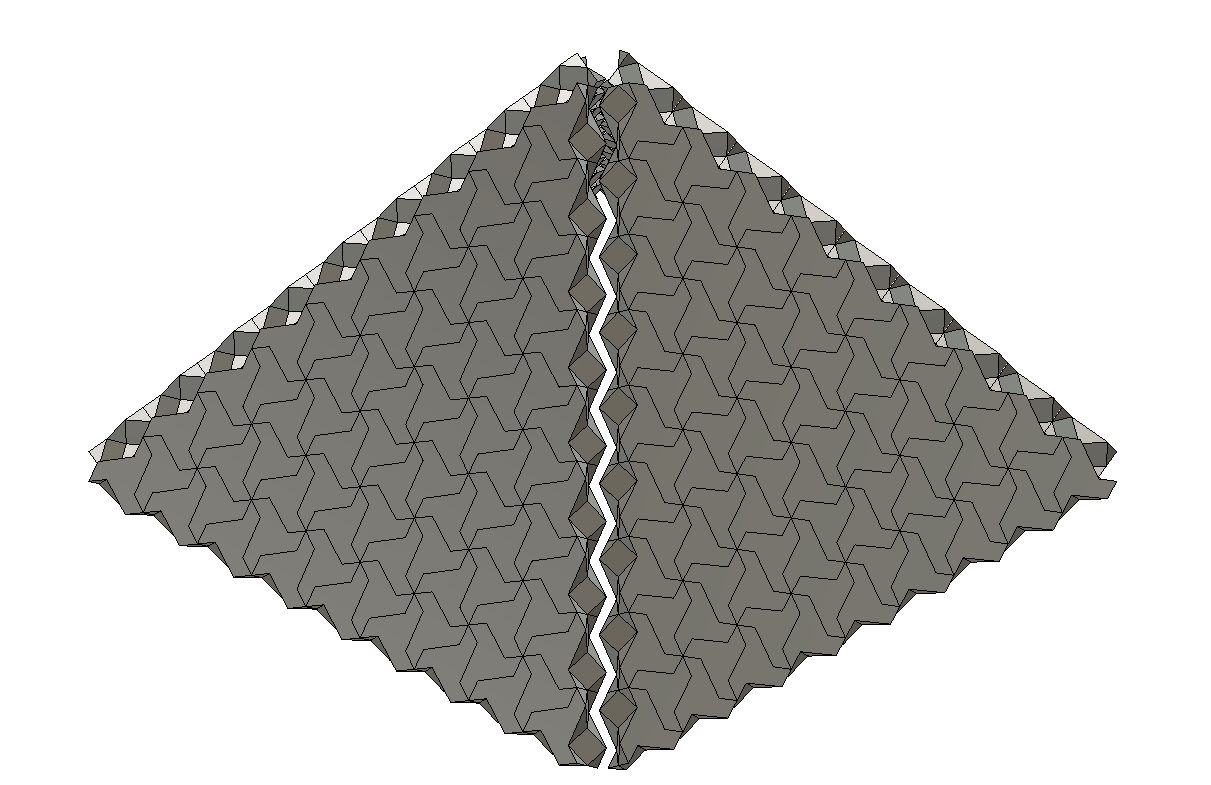}
    \end{subfigure}
    \hfill
    \caption{Two views of the blocks arranged to form the upper half of an octahedron}
    \label{fig:assembled-pyramid}
\end{figure}

The blocks described in Section \ref{ssec:individual-block} can be assembled to form a planar triangular face of a pyramid. Four identical flat pyramid faces can then be combined to construct the complete pyramid. As they are equilateral triangles, this assembly can be considered as the upper half of a regular octahedron. In the example presented here, the base of a face is composed of nine individual blocks and thus a face of the pyramid contains $81$ blocks in total. To join the faces, they are translated outward along the normals to their upper (outside) planes to ensure that no contact occurs between neighbouring faces. This procedure produces a pyramid-shaped assembly, where the upper and lower planes of the individual blocks form two pyramids, which we refer to as the upper (outside) and lower (inside) pyramid. The resulting construction is shown in Figure \ref{fig:assembled-pyramid}.

As the construction described so far does not yet enclose a volume, the remaining gaps must be filled. For this purpose, we introduce edge blocks, whose role is to fill the voids between the upper and lower faces along the edges where the pyramid’s faces meet. 

To avoid long elements running continuously along these edges we subdivide the edges as well. The subdivision is chosen so that the contact interfaces of the new modules do not coincide with the contacts between the adjacent blocks on each side. 

\begin{figure}[pos=ht]
    \centering
    \hfill
    \begin{subfigure}[b]{0.3\textwidth}
        \includegraphics[width=\textwidth]{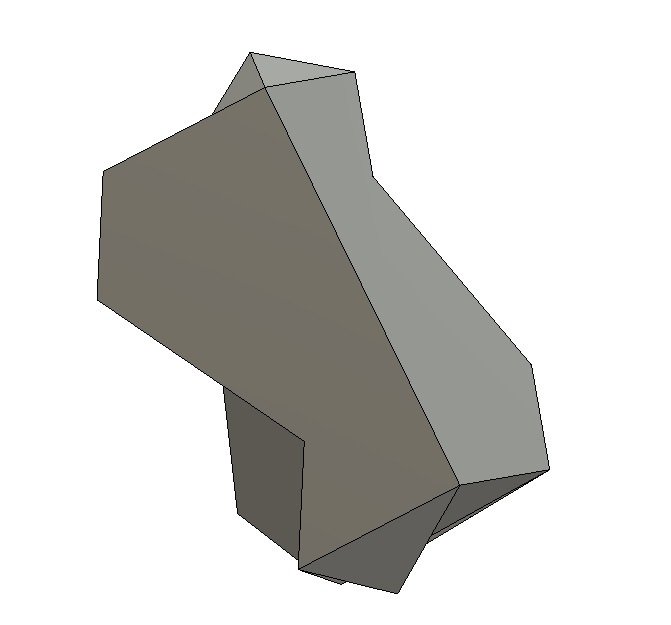}
    \end{subfigure}
    \hfill
    \begin{subfigure}[b]{0.3\textwidth}
        \includegraphics[width=\textwidth]{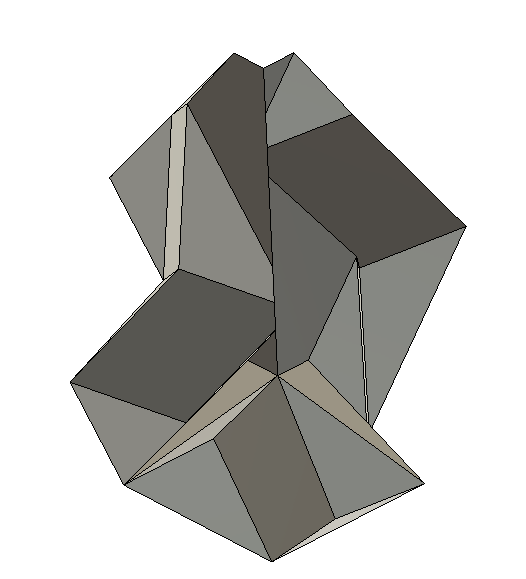}
    \end{subfigure}
    \hfill
    \begin{subfigure}[b]{0.3\textwidth}
        \includegraphics[width=\textwidth]{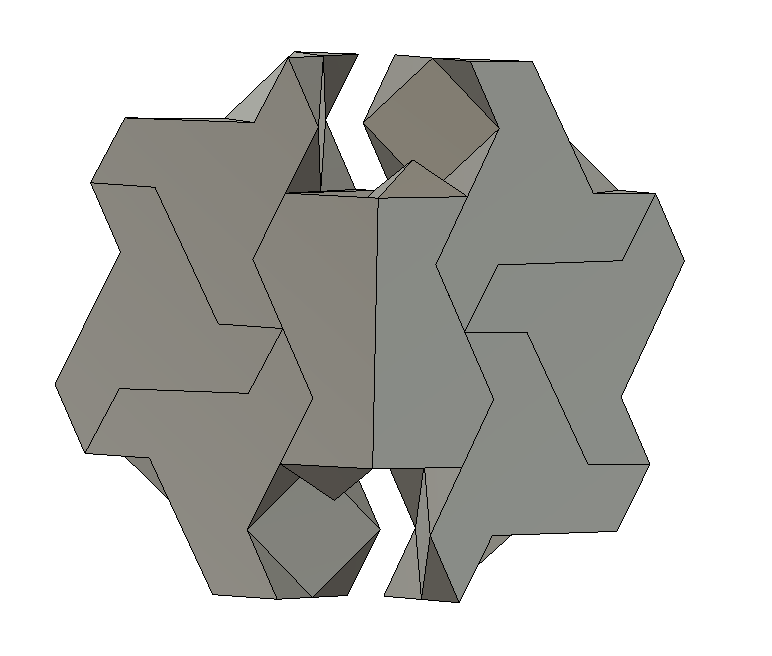}
    \end{subfigure}
    \hfill
    \caption{(a, b) Different views of a single edge block. (c) an edge block joining neighbouring blocks belonging to the two adjacent faces of the pyramid.}
    \label{fig:edge-block}
\end{figure}

The new interfaces of the edge blocks are then modified by appending or subtracting tetrahedral portions on their contact faces. This deformation is performed in such a way that the edge blocks have a rotational symmetry, which provides greater flexibility for assembly in practical implementations of this design. The resulting edge is shown in Figure \ref{fig:edge-block}.

\begin{figure}[pos=ht]
    \centering
    \hfill
    \begin{subfigure}[b]{0.4\textwidth}
        \includegraphics[width=\textwidth]{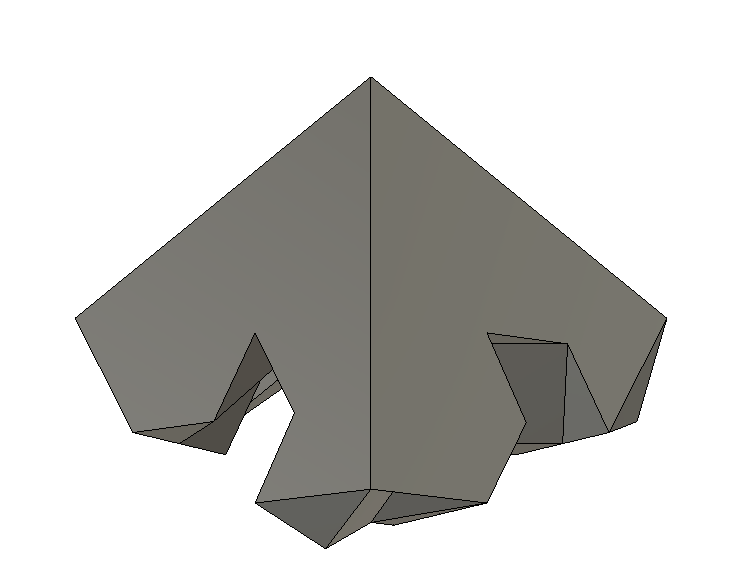}
    \end{subfigure}
    \hfill
    \begin{subfigure}[b]{0.4\textwidth}
        \includegraphics[width=\textwidth]{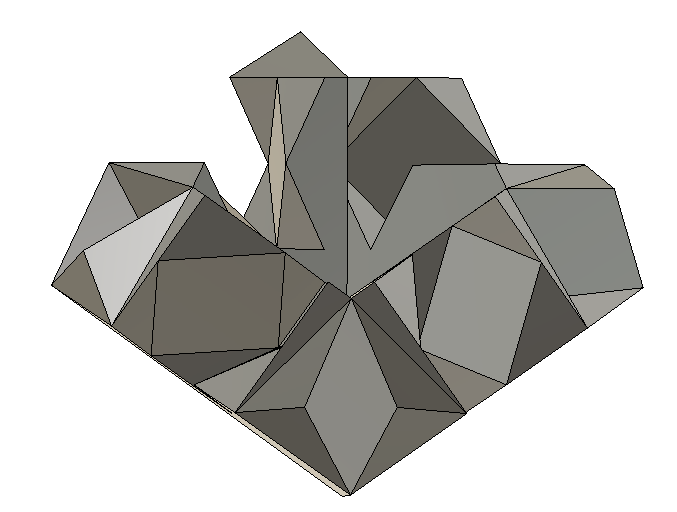}
    \end{subfigure}
    \hfill
    \caption{Different views of the apex block}
    \label{fig:apex-block}
\end{figure}

The last remaining element of the structure requiring consideration is the apex of the pyramid. Here the edge blocks overlap, and the remaining void must be fully filled without any intersections between neighbouring blocks. The resulting apex block is depicted in Figure \ref{fig:apex-block}.

\begin{figure}[pos=ht]
    \centering
    \hfill
    \begin{subfigure}[b]{0.3\textwidth}
        \includegraphics[width=\textwidth]{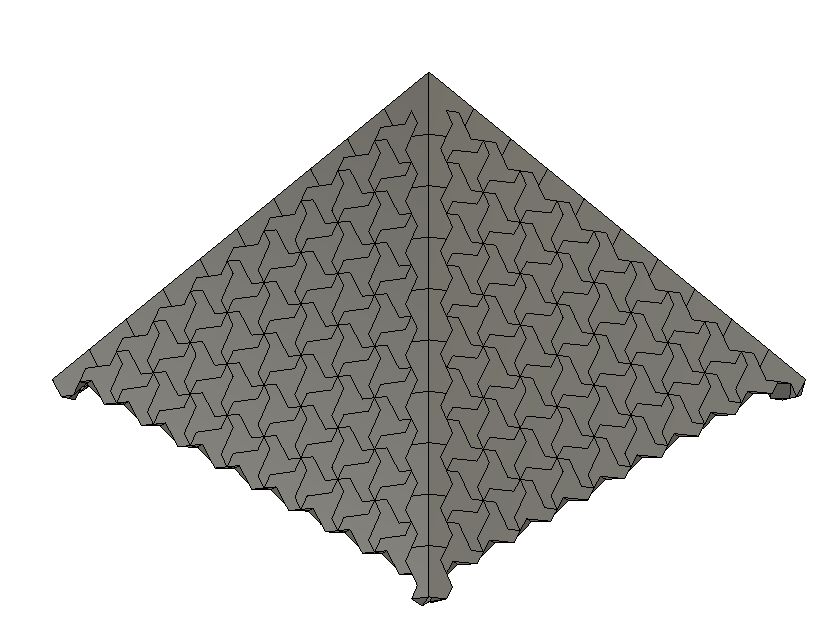}
        \caption{}
    \end{subfigure}
    \hfill
    \begin{subfigure}[b]{0.3\textwidth}
        \includegraphics[width=\textwidth]{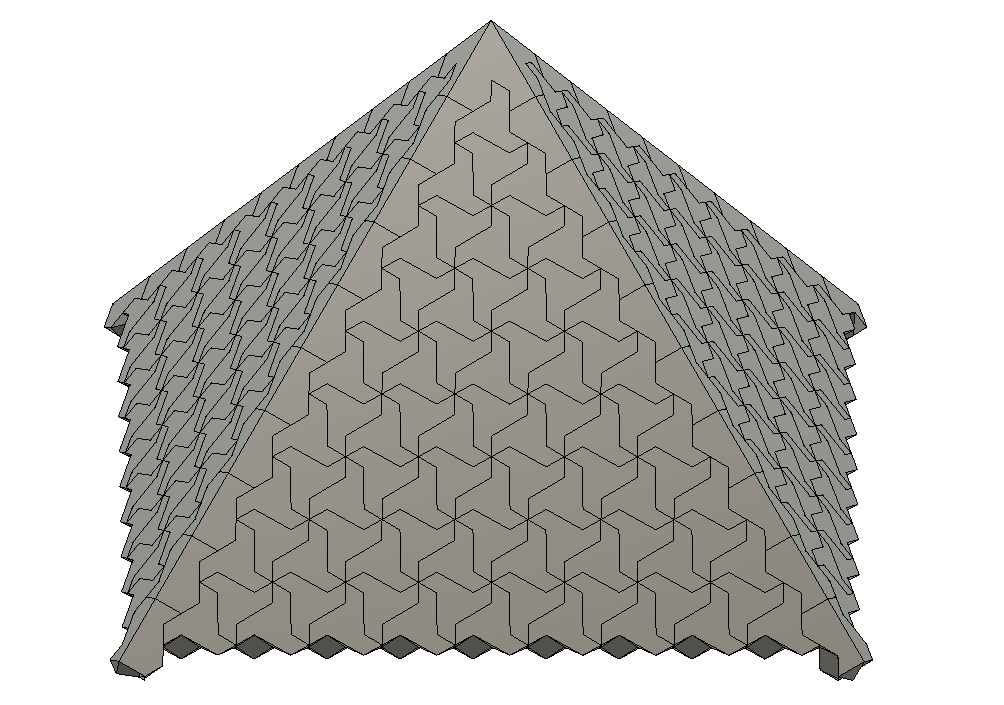}
        \caption{}
    \end{subfigure}
    \hfill
    \begin{subfigure}[b]{0.3\textwidth}
        \includegraphics[width=\textwidth]{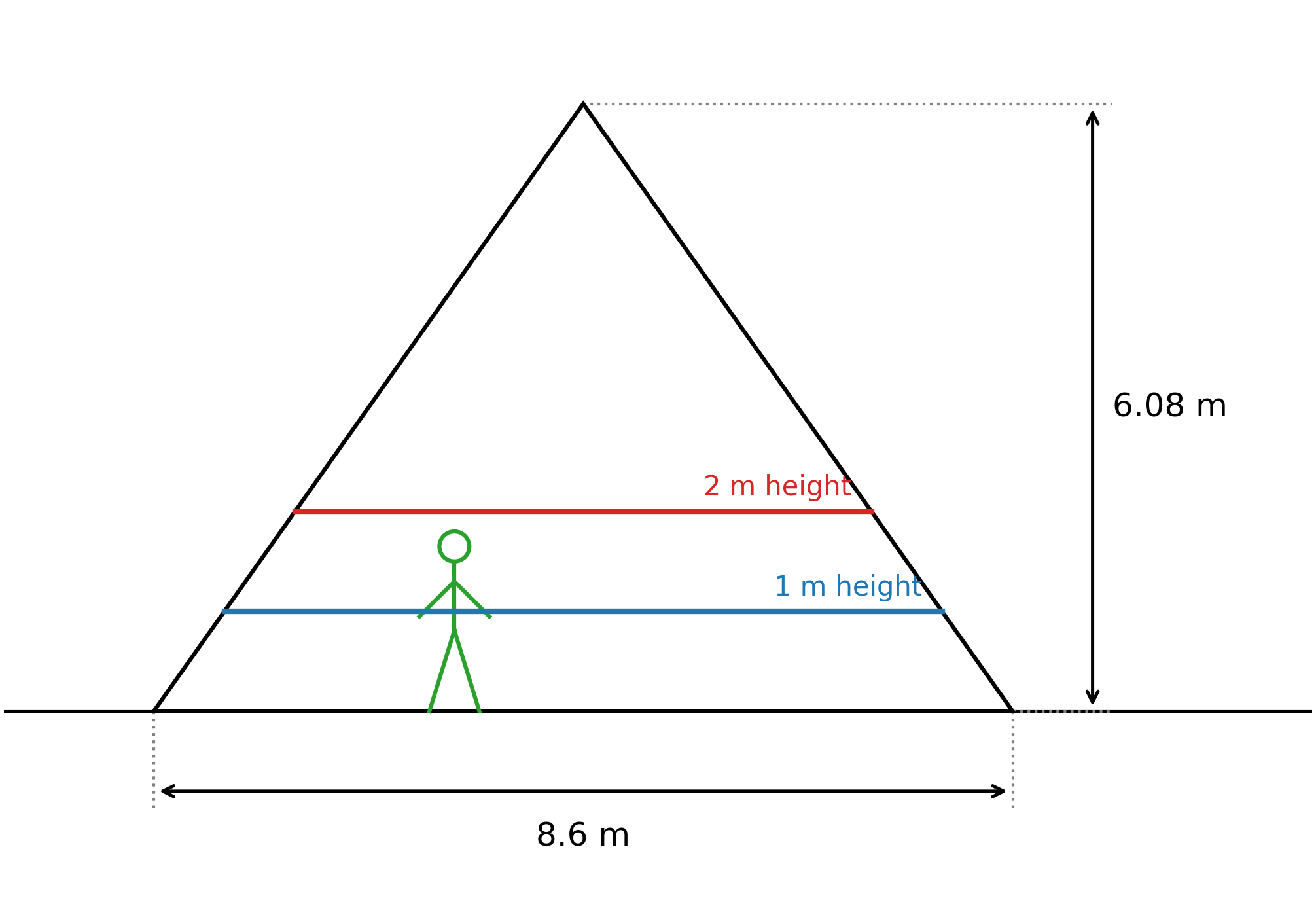}
        \caption{}
    \end{subfigure}
    \hfill
    \caption{Different views of the fully assembled pyramid (a, b), and cross section with lines for $1 \, \mathrm{m}$ and $2 \, \mathrm{m}$ markers.}
    \label{fig:pyramid-full}
\end{figure}

The complete structure thus generated approximates the square‑based pyramid we set out to construct, albeit with slightly modified edges along the bottom. The bottom of the pyramid is not perfectly planar;  its boundary follows the contour of the modified individual blocks. The remainder of the pyramid is fully planar and consists of four faces comprising $81$ interior blocks, $36$ edge blocks, and a single apex block. A picture of the full assembly is shown in Figure \ref{fig:pyramid-full}. 
The full assembly has an external base length of $\approx 8.6 \mathrm{m}$, and an external height of $\approx 6.1 \mathrm{m}$. This results in a usable volume of $\approx 198.87 \mathrm{m}^3$, assuming an internal base length of $7.5 \mathrm{m}$. Under the German legal framework for calculating living spaces \cite{woflv2003}, this corresponds to a living area of $\approx 29.19 \mathrm{m}^2$, of which $\approx 21.16 \mathrm{m}^2$ have full standing height (i.e., more than $2\, \mathrm{m}$ of clearance). If more space is required, the proposed geometry is inherently scalable, allowing larger habitats to be constructed using additional blocks of the same geometry. 

\section{Numerical analysis}\label{sec:numerical-analysis}
The individual sides of the pyramid form interlocks as discussed in \cite{schnelle2026} but this assumes that the peripheral blocks are fixed. In this case the peripheral blocks are the bottom row of blocks, the edge blocks, and the apex block. The bottom row can be fixed for example by anchoring it in the ground, but the edge blocks and the apex block remain unfixed in our proposed application. In what follows, we demonstrate that the resulting structure nevertheless maintains structural integrity. To examine the structural viability of the proposed assembly geometry, we ran multiple FE simulations with parameters aligned with what one may expect in lunar environment conditions.

\subsection{Simulation setup and assumptions}\label{ssec:setup-assumptions}
For the FE simulations we used Abaqus 2024 Explicit implementing Drucker-Prager plasticity \cite{dassault2024}. In general, a static solver would be more appropriate as the problem is fundamentally static. However, because the equilibrium is governed by a large number of inter‑block contacts, to improve convergence of the simulation, an explicit approach was chosen, despite its higher computational cost.

The simulations were run until $2 \, \mathrm{s}$ in the simulation passed, while all the loads were applied using a smooth step until $1 \, \mathrm{s}$. This ensured that the system reached an equilibrium, assessed by verifying that the internal energy substantially exceeded the kinetic energy. 

The assembly as described in Section 1 was constructed using the CAD software Fusion 360 \cite{autodesk2024} and exported as a single \verb|.step| file, which is also included in the data made available with this paper. This file contains all $361$ blocks making up the assembly. 

For the contact formulation, Abaqus \emph{general contact} was employed with a hard‑contact formulation and zero friction. This appears to be admissible in a first approximation, since for topological interlocking systems, the structural integrity is enabled primarily by the block geometry and arrangement. Friction between blocks does play a role \cite{djumas2016}, but has a lesser effect.

\subsection{Lunar environment}\label{ssec:lunar-env}
For more extensive literature on the lunar environment we direct the reader to \cite{benaroya2017, sourcebook1991}. 
For our application the most relevant factors are the gravity on the lunar surface as well as the almost completely missing atmosphere. Therefore, for the simulations we applied a gravitational load of around $\frac{1}{6} \mathrm{g}$ (where $\mathrm{g}$ is the gravitational acceleration at the Earth’s surface), specifically $1.62 \frac{\mathrm{m}}{\mathrm{s}^2}$. As the atmosphere on the lunar surface has a very low density ( $\approx 10^4 \frac{\mathrm{molecules}}{\mathrm{cm}^3}$ compared to $\approx2.5 \, 10^{19} \frac{\mathrm{molecules}}{\mathrm{cm}^3}$ on Earth), the simulation included no external atmospheric loads. However, to provide conditions humans can sustain for prolonged periods of time, an internal positive pressure needs to be maintained. We therefore adopt a conservative internal‑pressure specification based on the cabin environment of commercial airliners during cruise, approximately $0.8\, \mathrm{atm}$ \cite{kelly2007}. This corresponds to an internal pressure of about $81,060 \, \mathrm{Pa}$, which was applied to all surfaces bounding the pressurised interior. 
A main concern in the context of lunar habitat is that the low lunar atmosphere offers very limited protection against cosmic radiation. We do not explicitly address radiation shielding in this work, but for structural integrity we do need an additional overburden on the outer surface. A layer of loose regolith may serve both needs: structural stabilisation and radiation shielding. We therefore propose to add such a layer that exerts an external pressure of $0.15 \, \frac{\mathrm{N}}{\mathrm{mm}^2}$ on the habitat surface. We do not evaluate the level of radiation protection this layer would provide. Further radiation protection measures may be added if required \cite{bell2011}.

\subsection{Material and damage model}\label{ssec:mat-model}
Since the habitat is proposed to be constructed from sintered lunar regolith, the material properties used in the FE simulations were chosen to approximate the best‑available data for such materials. The model was assigned a Young’s modulus of $15 \, \mathrm{GPa}$ and a Poisson ratio of $0.17$, values representative of published measurements for sintered regolith simulants \cite{wang2025, zhang2023}. These parameters place the material within the range of porous ceramics \cite{ashby1993}. To account for gravitational loading, the material density is also required. We adopt a value of $1.5 \, \frac{\mathrm{g}}{\mathrm{cm}^3}$, consistent with a lightweight, partially densified regolith‑derived ceramic \cite{gao2024}. This choice provides a reasonable approximation of sintered regolith while remaining conservative for structural analysis. 
The governing failure mode for the habitat is expected to be a brittle fracture of the sintered regolith \cite{zhu2026}. To capture this behaviour in the simulations, we employ the \emph{Drucker-Prager} model, which provides a convenient phenomenological framework for pressure‑dependent yielding. We adapt it for brittle behaviour by identifying regions within the blocks within the assembly that enter the plastic strain regime and treating those regions as failed. 

The model parameters were chosen to reflect available data for sintered regolith and related granular-ceramic materials. The friction angle was set to $58.5^\circ$ \cite{yang2014}, the flow stress ratio to $1$ (assuming isotropy), and the dilation angle to $0^\circ$, consistent with a non‑dilatant brittle material. The tensile yield stress was taken as $5\, \mathrm{MPa}$ \cite[Tab. 7]{zhou2024}.

To determine failure of the assembly, we consider two failure modes of the proposed habitat. The first one is the loss of global structural integrity. If, for example, the internal pressure is so high that the blocks are displaced outward under the expected lunar environment conditions, the habitat is deemed structurally unsound. The second failure mode is local failure of individual blocks. As these blocks are made from a brittle material, they may fracture and spall. To determine if a block with such a broken off portion is still providing the interlocking capability essential for structural integrity, we employ a Python script that maps all failed regions across all blocks. If the remaining portion of a block remains intact, such a local failure is considered to be acceptable. This is consistent with the results of \cite{schnelle2026}, where it was shown that for the Starfish geometry, the degree of deformation affects the mechanical response of an assembly, yet its load‑bearing capacity is preserved across the entire deformation range. 

\section{Results}\label{sec:results}
We ran four simulations to investigate the mechanical response of the proposed habitat system under the loads expected in the lunar environment. The first simulation used a fully intact habitat, exactly as described in Section \ref{sec:geom}. 

To investigate the mechanical response of the habitat when a block is missing, we simulated three additional assemblies identical to that described in Section \ref{sec:geom}, each with one block removed. Because the structure is symmetric, all missing‑block locations were placed on the same face of the pyramid. The removed blocks were positioned near the base, at mid‑height, and close to the apex, covering a range of plausible failure locations. 

After the simulations were run, we first verified that the assembly had reached equilibrium. This was done by comparing the internal energy with the kinetic energy over the simulation time. After all loads were fully applied at $1\, \mathrm{s}$, the two energies levelled off and remained essentially constant for the remainder of the simulation. This behaviour indicates that the assembly carries the applied loads through force transfer between the blocks rather than through continued motion. 

Next, we evaluated the displacement of the material points of the pyramid faces in the Y‑direction, aligned with the direction of gravitational loading. For the full assembly, the maximum absolute deformation was $-0.071 \, \mathrm{mm}$. For the assembly with a missing block at the bottom‑level, mid-height, and near-apex level, the corresponding values were $-1.56 \, \mathrm{mm}$, $-1.61 \, \mathrm{mm}$ and $-1.60 \, \mathrm{mm}$, respectively. 

\begin{figure}[pos=ht]
    \centering
    \hfill
    \begin{subfigure}[b]{0.4\textwidth}
        \includegraphics[width=\textwidth]{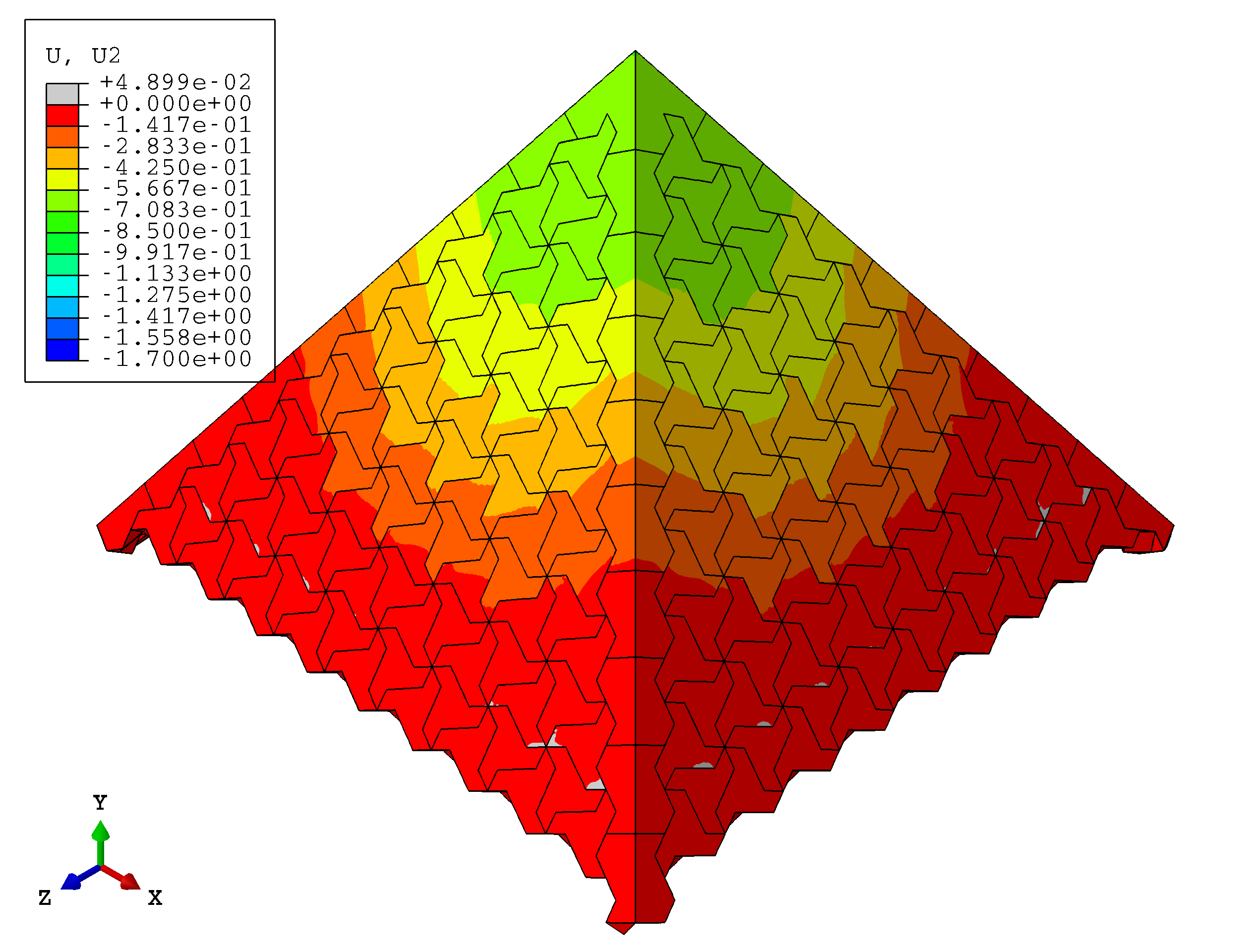}
    \end{subfigure}
    \hfill
    \begin{subfigure}[b]{0.4\textwidth}
        \includegraphics[width=\textwidth]{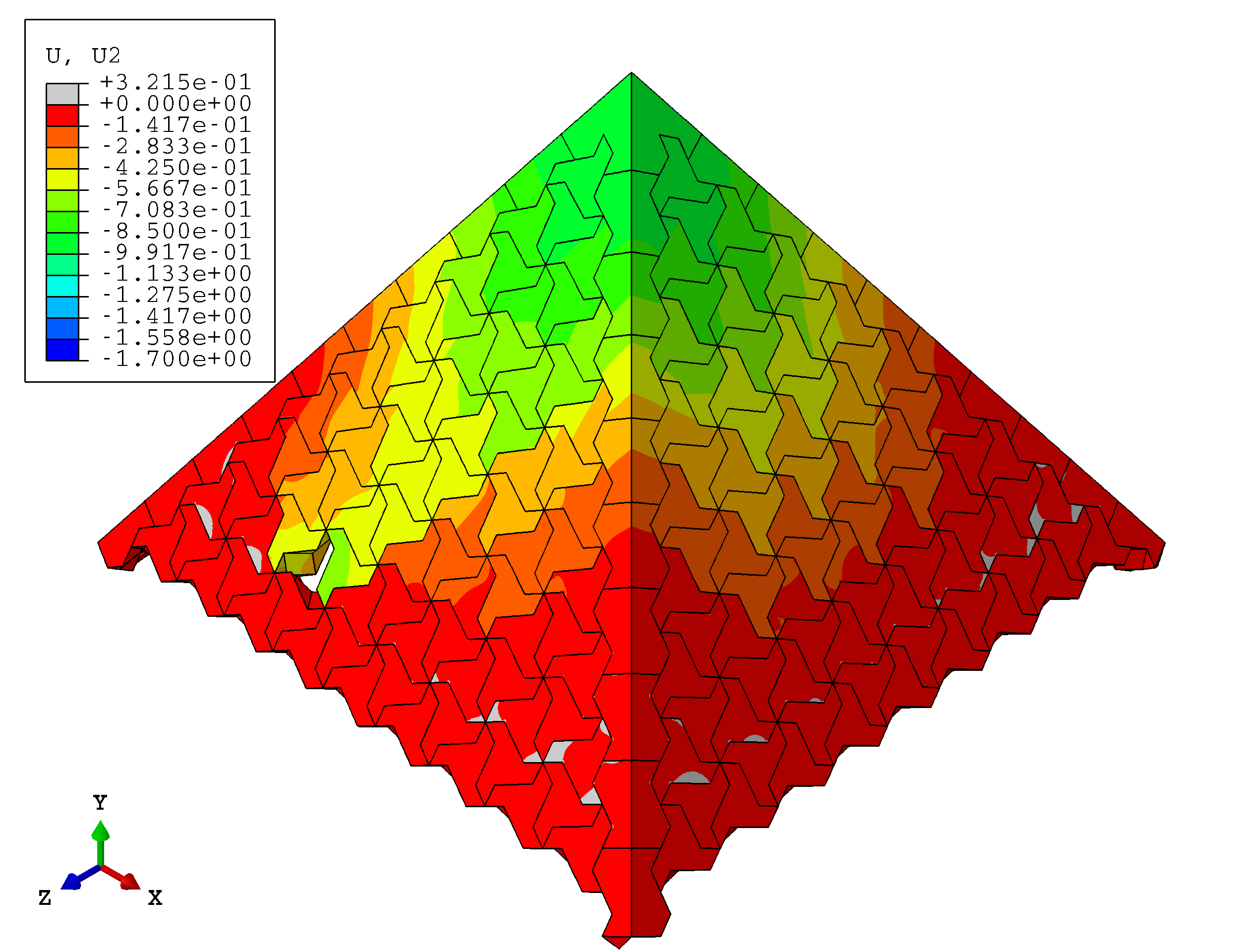}
    \end{subfigure}
    \hfill
    \\
    \hfill
    \begin{subfigure}[b]{0.4\textwidth}
        \includegraphics[width=\textwidth]{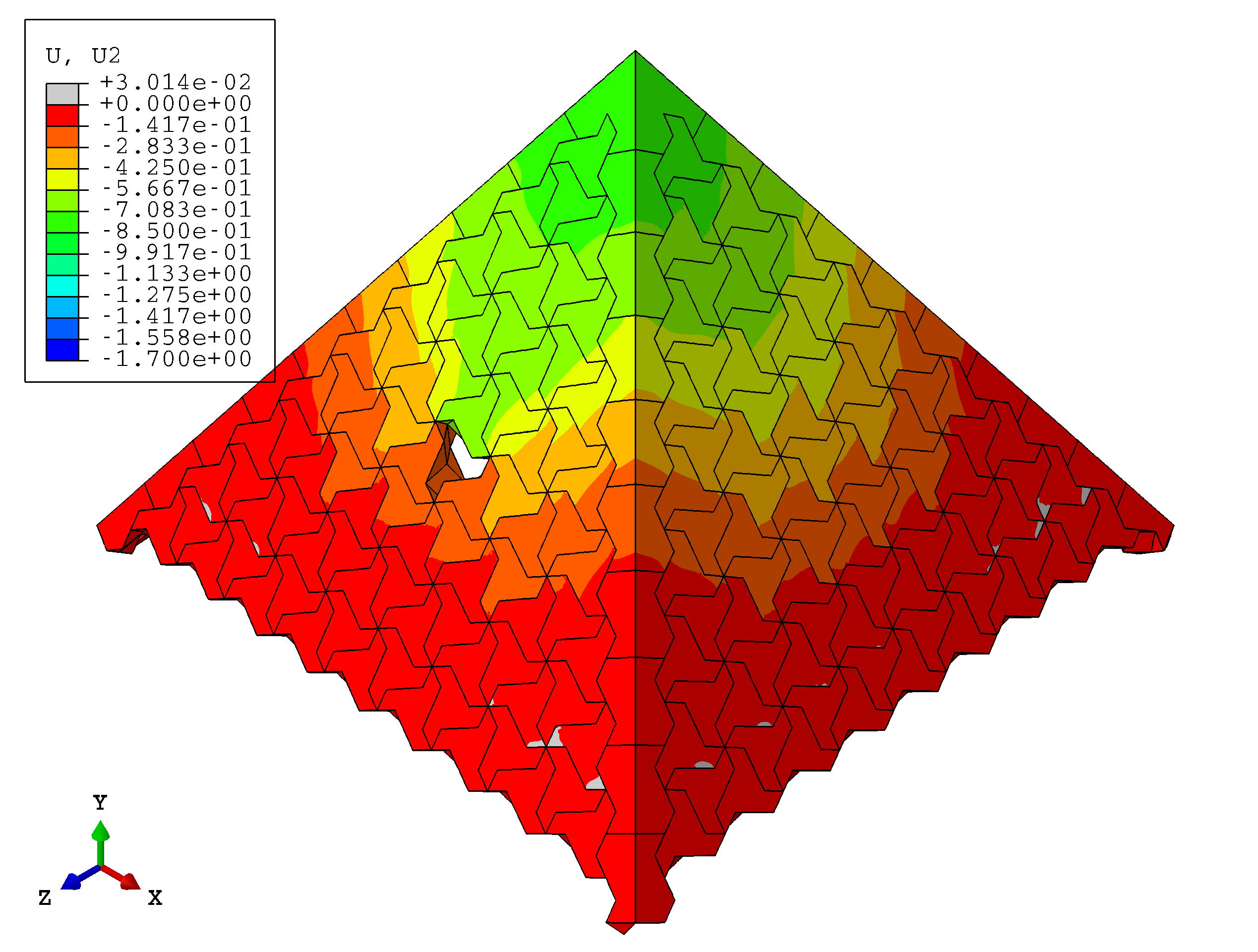}
    \end{subfigure}
    \hfill
    \begin{subfigure}[b]{0.4\textwidth}
        \includegraphics[width=\textwidth]{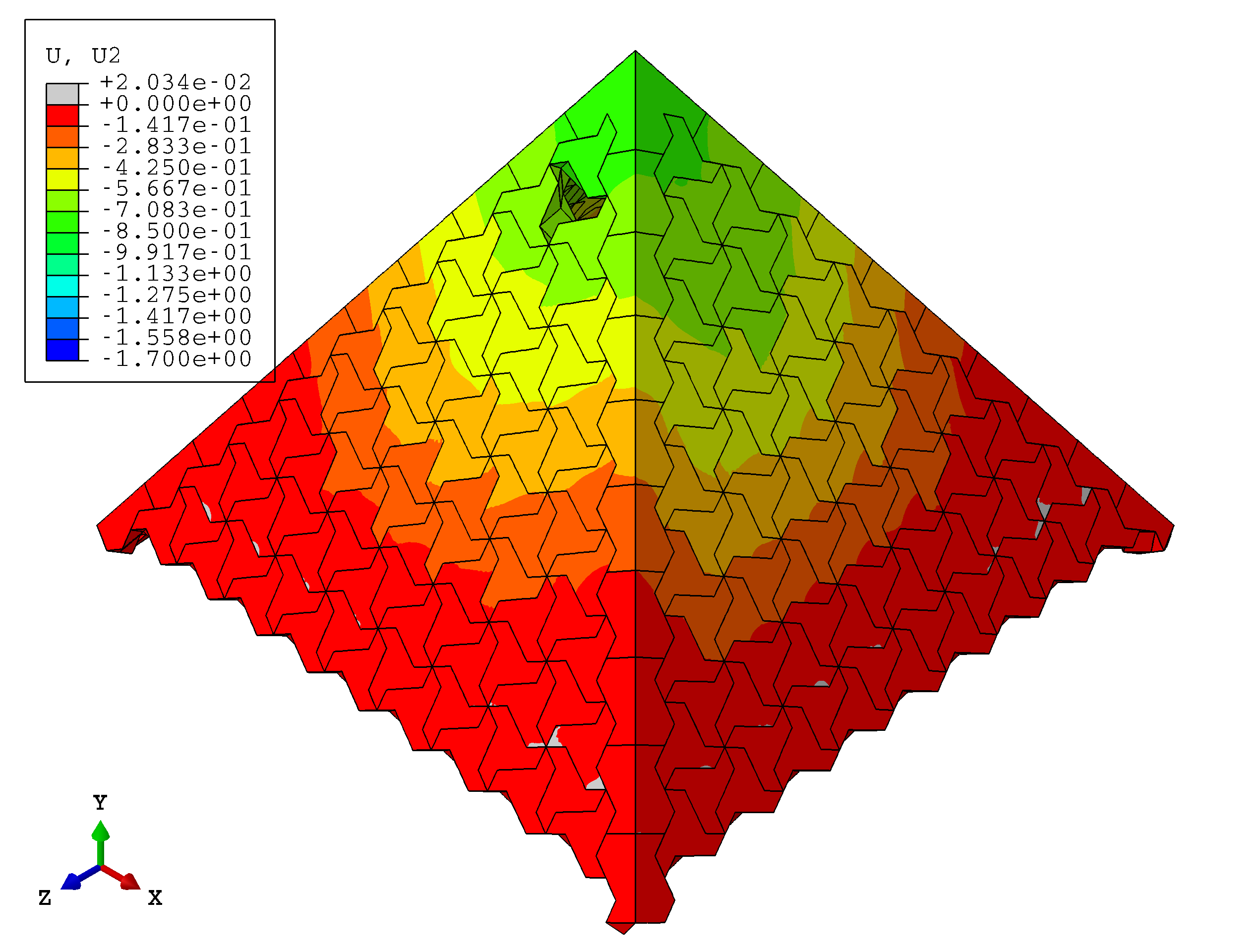}
    \end{subfigure}
    \hfill
    \caption{The total deformation for the four tested habitat variants: (a) no missing blocks, (b) a missing block near the bottom, (c) a missing block at mid‑height, and (d) a missing block near the apex. All displacements are shown using a uniform scale in millimetres.}
    \label{fig:U2-FEM}
\end{figure}

\begin{figure}[pos=ht]
    \centering
    \hfill
    \begin{subfigure}[b]{0.4\textwidth}
        \includegraphics[width=\textwidth]{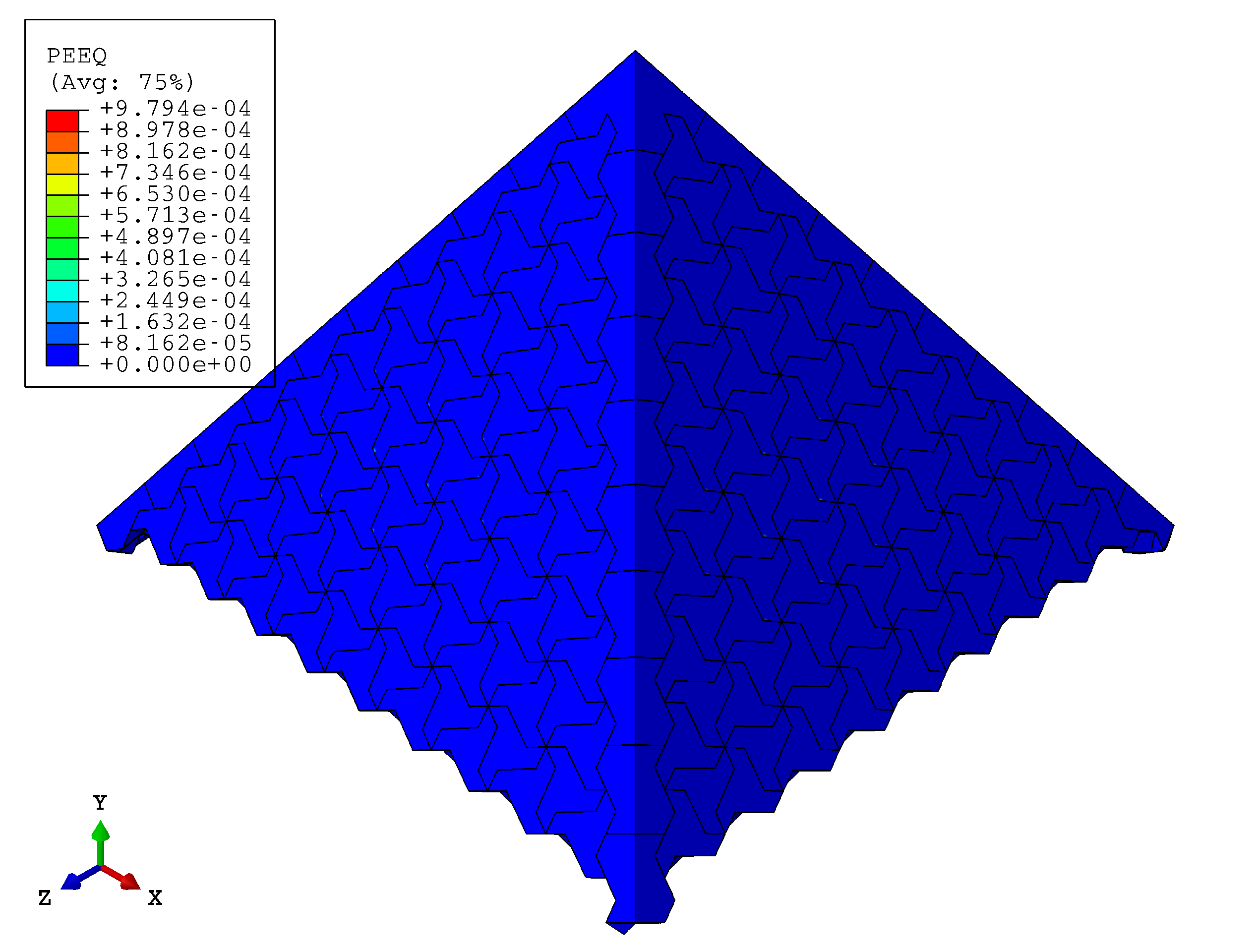}
    \end{subfigure}
    \hfill
    \begin{subfigure}[b]{0.4\textwidth}
        \includegraphics[width=\textwidth]{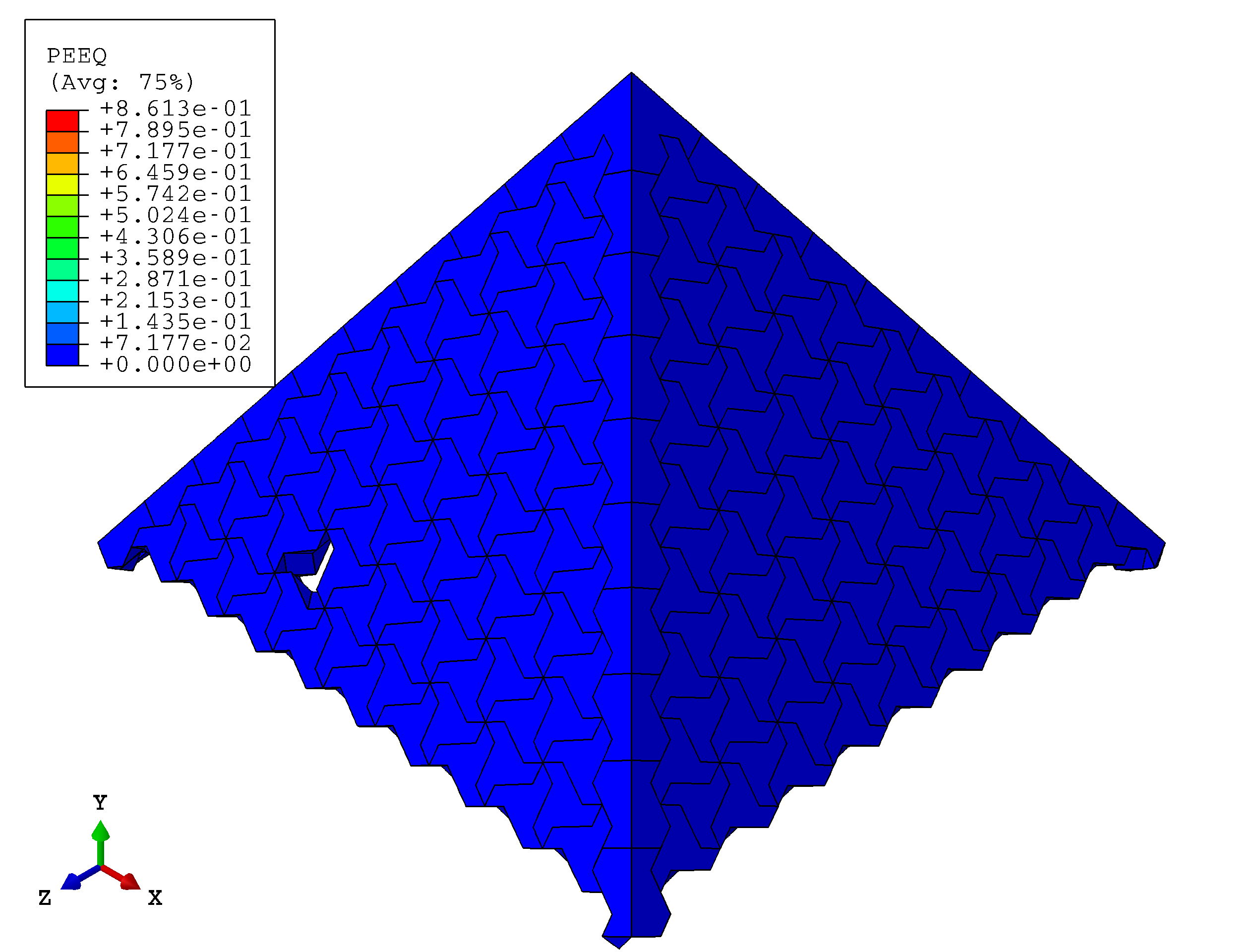}
    \end{subfigure}
    \hfill
    \\
    \hfill
    \begin{subfigure}[b]{0.4\textwidth}
        \includegraphics[width=\textwidth]{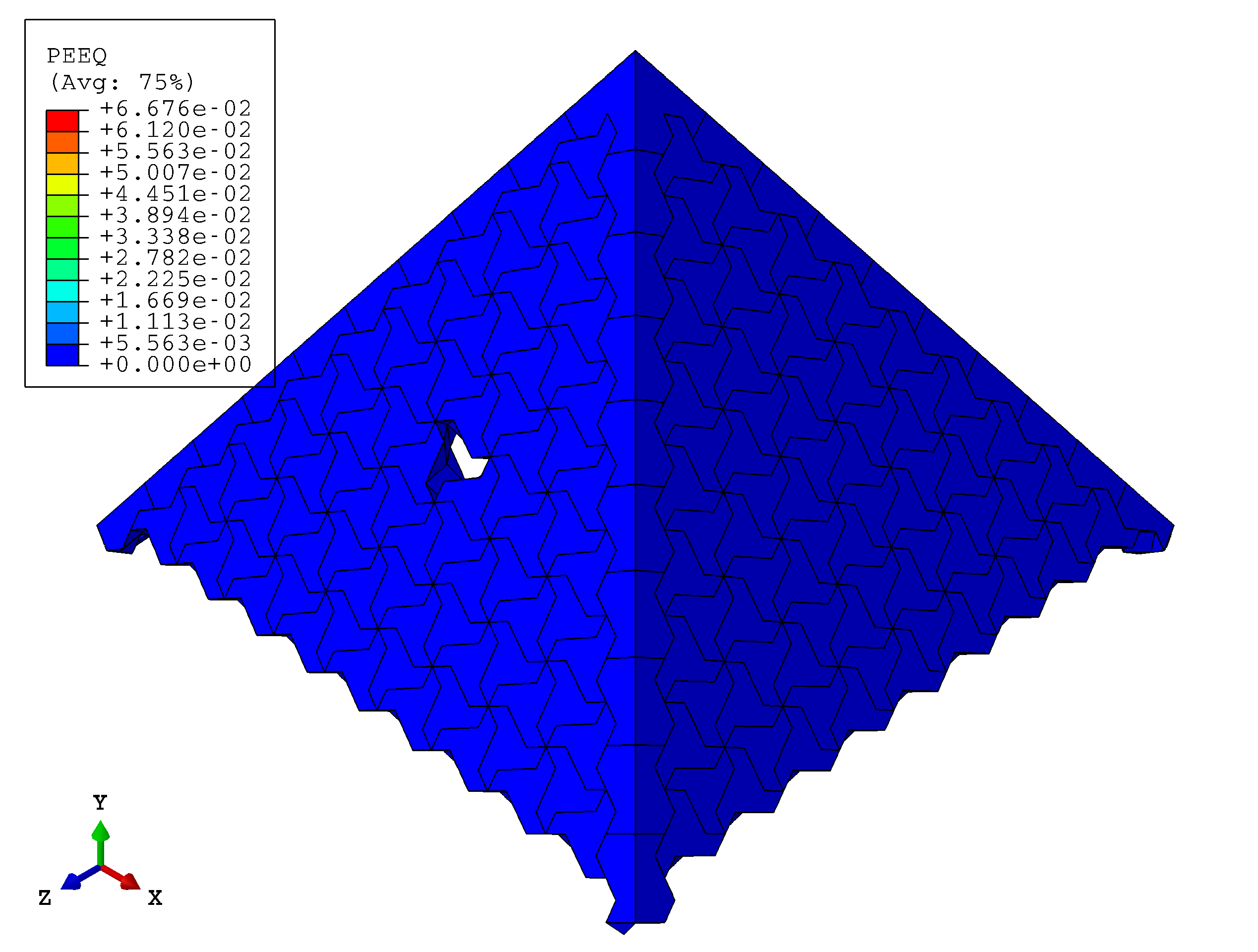}
    \end{subfigure}
    \hfill
    \begin{subfigure}[b]{0.4\textwidth}
        \includegraphics[width=\textwidth]{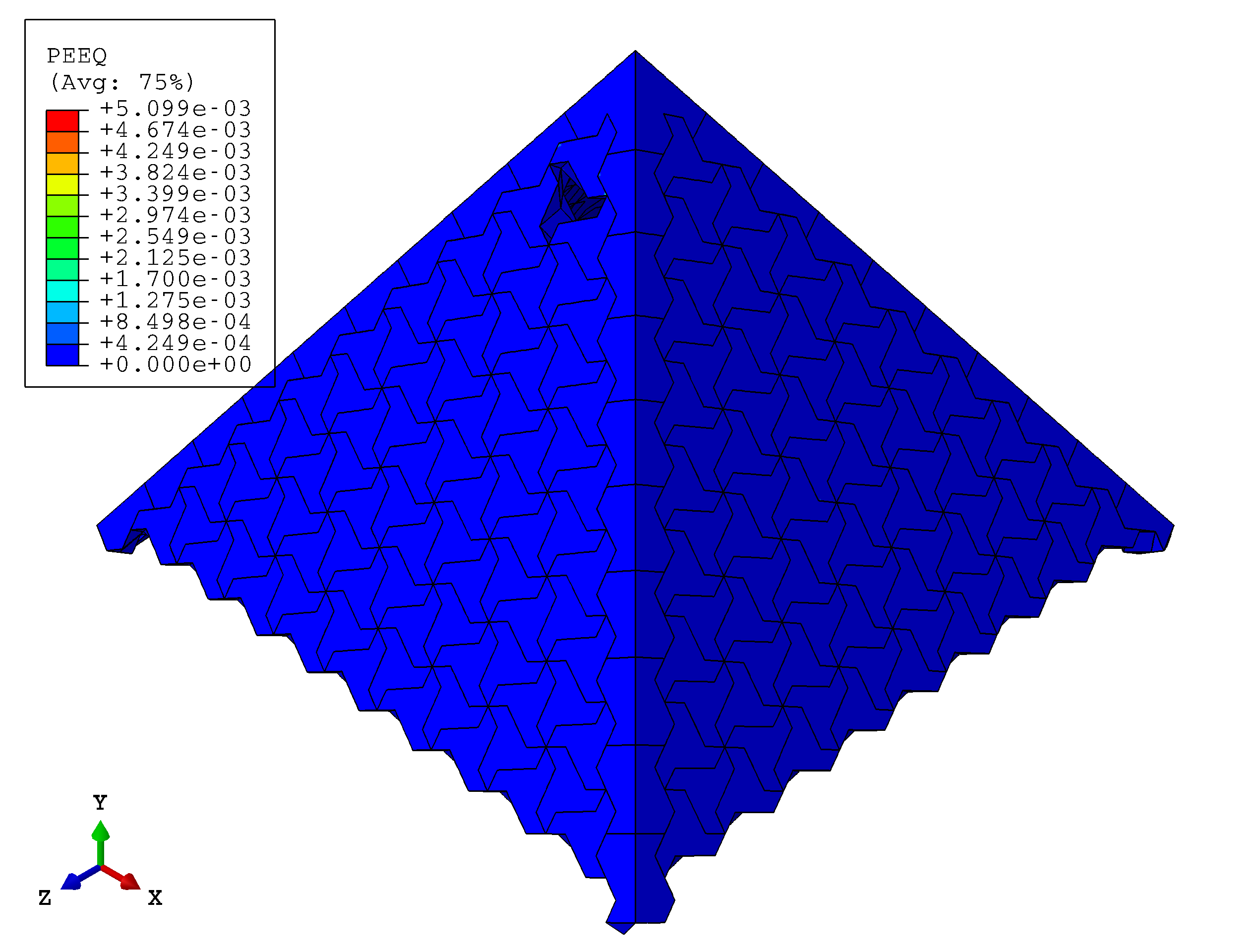}
    \end{subfigure}
    \hfill
    \caption{PEEQ values for the four tested habitat variants: (a) no missing blocks, (b) a missing block near the bottom, (c) a missing block at mid‑height, and (d) a missing block near the apex.}
    \label{fig:PEEQ-FEM}
\end{figure}

Next we evaluate the local failure criterion in terms of the location of any region of a block where the principal equivalent stress exceeds the tensile yield stress, set at $5 \, \mathrm{MPa}$. In the global view shown in Figure \ref{fig:PEEQ-FEM}, all variants appear structurally acceptable, as these values throughout the pyramid remain below the prescribed tensile yield stress. However, zooming in on the regions where the principal equivalent plastic strain (PEEQ) attains its maximum values reveals small areas where it is non-zero. These instances of plastic yielding are confined to very limited, localised portions of individual blocks. The key question, therefore, is whether the loss of such small regions - if they were to be considered effectively removed - would compromise the topological interlocking behaviour of the assembly. In other words, we examine whether this localised damage could disrupt the geometric constraints that enable load transfer and structural robustness in the interlocked system. 

\begin{figure}[pos=ht]
    \centering
    \hfill
    \begin{subfigure}[b]{0.4\textwidth}
        \includegraphics[width=\textwidth]{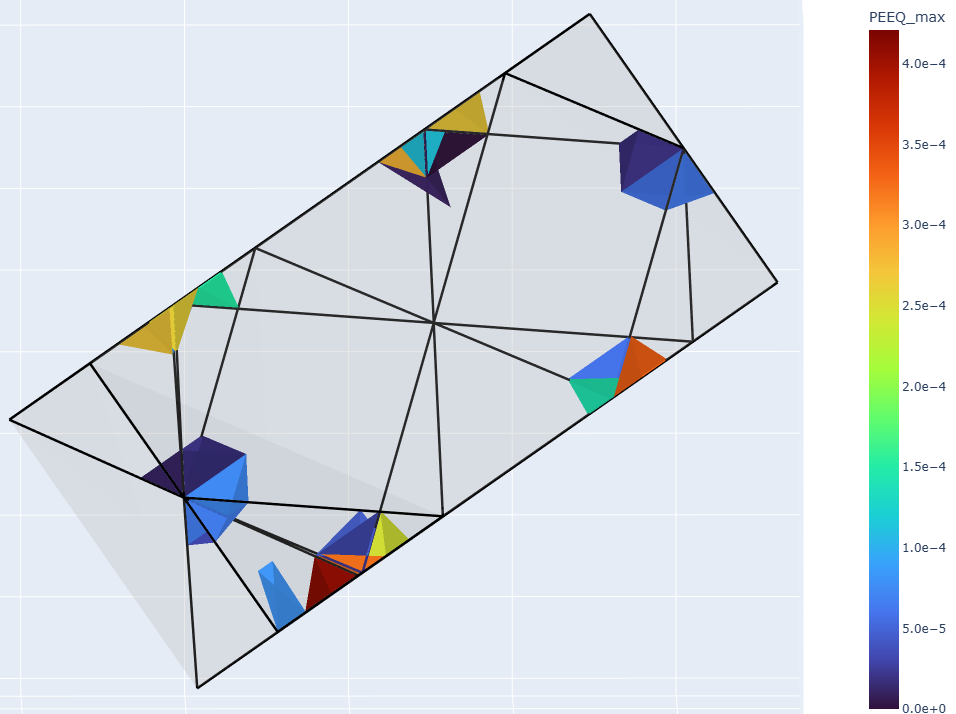}
    \end{subfigure}
    \hfill
    \begin{subfigure}[b]{0.4\textwidth}
        \includegraphics[width=\textwidth]{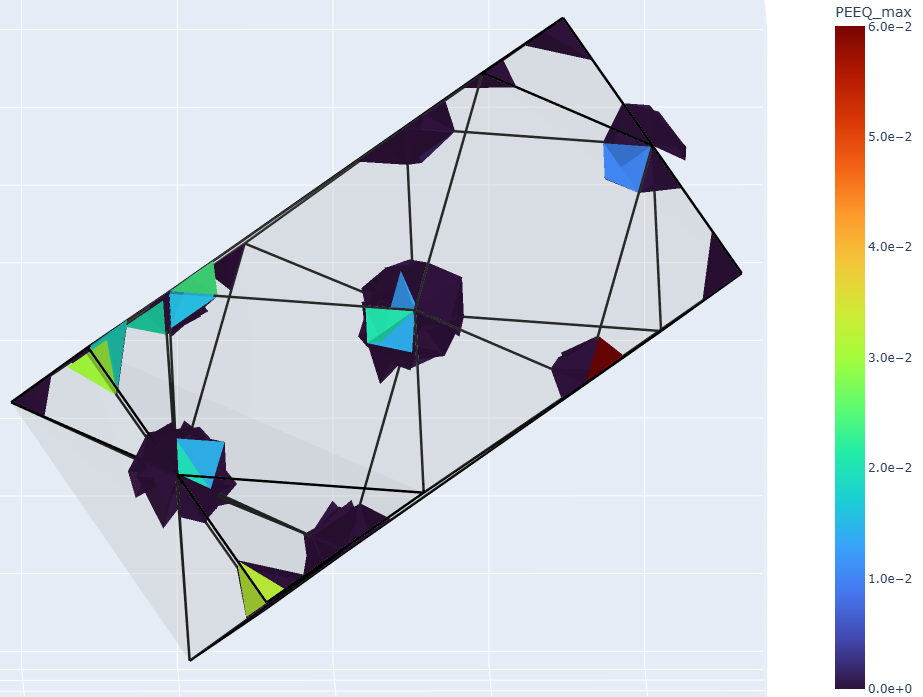}
    \end{subfigure}
    \hfill
    \\
    \hfill
    \begin{subfigure}[b]{0.4\textwidth}
        \includegraphics[width=\textwidth]{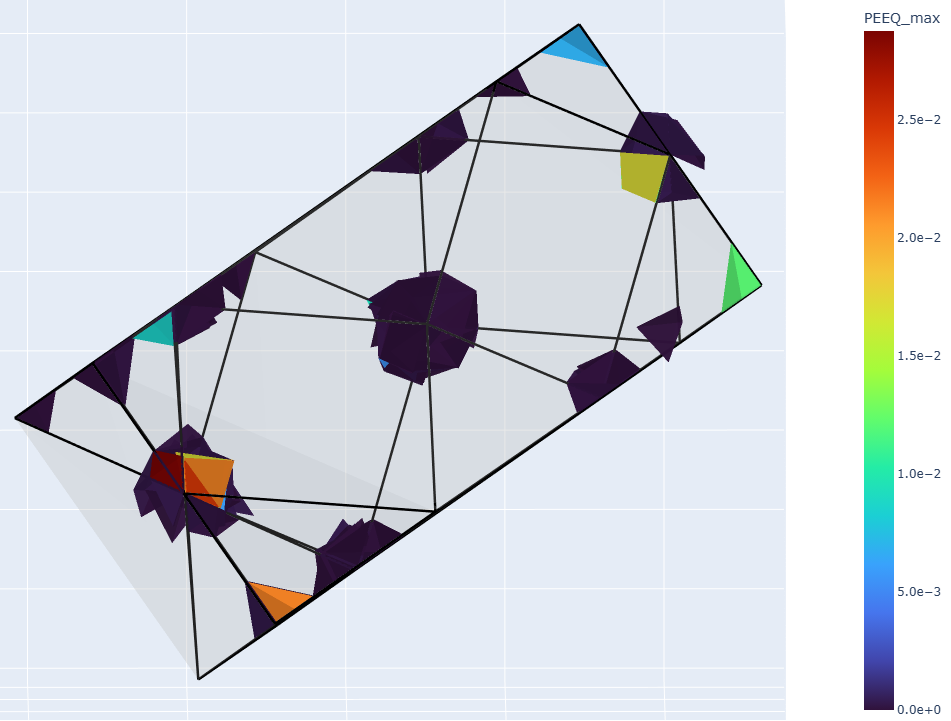}
    \end{subfigure}
    \hfill
    \begin{subfigure}[b]{0.4\textwidth}
        \includegraphics[width=\textwidth]{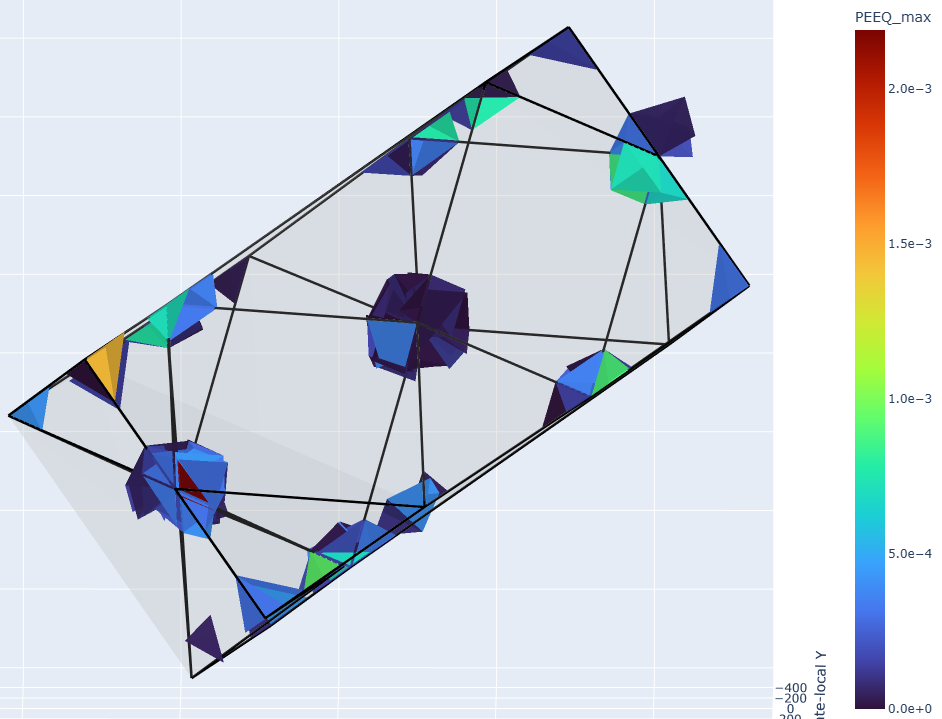}
    \end{subfigure}
    \hfill
    \caption{Side views of the mapped PEEQ values for all double-Starfish blocks in the four tested variants: (a) no missing blocks, (b) a missing block near the bottom, (c) a missing block at mid‑height, and (d) a missing block near the apex. }
    \label{fig:superimposed-FEM}
\end{figure}

First of all we note that non‑zero PEEQ values occur only in the double‑Starfish blocks, and therefore restrict our analysis to this part of the assembly. To carry out this analysis, we used two Python scripts, both of which are provided in the data accompanying this work. The first script exported the PEEQ values from all Abaqus simulations to a \verb|.csv| file together with their spatial coordinates. This enabled the second script to maps all double‑Starfish blocks occurring across the simulations. The result is an outline of the double‑Starfish block showing all locations where the PEEQ exceeds $1 \, 10^{-7}$, in accordance with the threshold defined in this work. Side views for all four configurations are shown in Figure \ref{fig:superimposed-FEM}. 

Figure \ref{fig:superimposed-FEM} shows that the regions exhibiting plastic yielding all lie on three planes: the two side planes facing the interior and exterior of the habitat, and the middle plane corresponding to the transition between the two deformation modes of the fundamental domain. The remaining bulk of the blocks situated between these planes remains fully intact. Consequently, the topological interlocking - central to the structural integrity of the pyramid - remains preserved. 
Figure \ref{fig:superimposed-FEM} also makes it clear that both the magnitude of plastic strain and the spatial extent of the affected regions are greater in all three configurations with a missing block. This observation is consistent with the higher maximum PEEQ values in Figure \ref{fig:PEEQ-FEM} for the missing‑block simulations, and likewise with the larger maximum deformations observed in Figure \ref{fig:U2-FEM} for these cases.

\section{Discussion}\label{sec:discussion}
In the present work, we first derived a geometry that satisfies the requirements set out at the outset: a modular structure based on a topologically interlocked assembly composed of three distinct block types, capable of being expanded, and enclosing a usable internal volume. This geometry was then evaluated under the expected lunar environmental conditions, both in the complete assembly and in three configurations where one block was removed. We demonstrated that a habitat based on topological interlocking is feasible and that the structure retains a significant degree of integrity even when individual blocks are missing. This confirms one of the principal advantages of topological interlocking: inherent tolerance of TI structures to localised failure or loss of a module. 

In addition, the proposed construction method enables a more holistic lightweighting strategy through in‑situ resource utilisation (ISRU). Current approaches are often constrained by the size of the habitat to be constructed on the Moon or require the entire structure to be delivered from Earth, resulting in substantial mass penalties. By contrast, the present approach allows the bulk of the structural material to be sourced directly on the lunar surface.

Several assumptions underlie the present work, the most significant concerning the material properties of lunar regolith. Although terrestrial analogue materials have been shown in the literature to be sinterable, the resulting material properties may differ substantially from those of actual lunar regolith. 

In addition, the habitat proposed in this work does not yet incorporate provisions for entry and exit, and a dedicated system for this purpose might need to be transported from Earth. Such a system might be integrated within a TIA‑based design, but this lies outside the scope of the present study.  

Furthermore, we assume the presence of a membrane to ensure gas‑tightness. This design measure is necessary because the geometry of TIAs, with gaps between the blocks, does not provide a hermetic enclosure of the internal volume, and these gaps may further change under load. The authors are not aware of existing research on the gas‑tightness of TIA‑based structures, and this aspect of habitat design will need to be addressed in future work.  

Lastly, the proposed habitat is based on an existing geometry with only minor modifications. TIAs possess a large number of parameters that determine the exact block geometry, and a purpose‑optimised design, tailored entirely to the requirements of a lunar habitat, might perform even better.

In summary, this work demonstrates numerically that it is feasible to construct a habitat based on TIAs, which - given current understanding of the available material - can be realised largely through ISRU. The design accommodates the inherent brittleness of sintered regolith and enables the internal pressurisation loads to be carried effectively as well as future reuse of the constructed blocks through expansion.


\printcredits\\
The authors gratefully acknowledges the support by the Deutsche Forschungsgemeinschaft (DFG, German Research Foundation) in the framework of the Collaborative Research Centre CRC/TRR 280 “Design Strategies for Material-Minimized Carbon Reinforced Concrete Structures – Principles of a New Approach to Construction” (project ID 417002380).

\bibliographystyle{unsrtnat}

\bibliography{refs}



\end{document}